\documentstyle[12pt]{article}
\begin{document}    
%%%%%%%%%%%%%%%%%%%%% YITP/U defs. %%%%%%%%%%%%%%%%%%%%%%%%
\ifx\TwoupWrites\UnDeFiNeD\else\target{\magstepminus1}{11.3in}{8.27in}
	\source{\magstep0}{7.5in}{11.69in}\fi
\newfont{\fourteencp}{cmcsc10 scaled\magstep2}
\newfont{\titlefont}{cmbx10 scaled\magstep2}
\newfont{\authorfont}{cmcsc10 scaled\magstep1}
\newfont{\fourteenmib}{cmmib10 scaled\magstep2}
	\skewchar\fourteenmib='177
\newfont{\elevenmib}{cmmib10 scaled\magstephalf}
	\skewchar\elevenmib='177
\newif\ifpUbblock  \pUbblocktrue
\newcommand\nopubblock{\pUbblockfalse}
\newcommand\topspace{\hrule height 0pt depth 0pt \vskip}
\newcommand\pUbblock{\begingroup \tabskip=\hsize minus \hsize
	\baselineskip=1.5\ht\strutbox \topspace-2\baselineskip
	\halign to\hsize{\strut ##\hfil\tabskip=0pt\crcr
	\the\Pubnum\crcr\the\date\crcr}\endgroup}
\newcommand\YITPmark{\hbox{\fourteenmib YITP\hskip0.2cm
        \elevenmib Uji\hskip0.15cm Research\hskip0.15cm Center\hfill}}
\renewcommand\titlepage{\ifx\TwoupWrites\UnDeFiNeD\null\vspace{-1.7cm}\fi
	\YITPmark\vskip0.6cm
	\ifpUbblock\pUbblock \else\hrule height 0pt \relax \fi}
\newtoks\date
\newtoks\Pubnum
\newtoks\pubnum
\Pubnum={YITP/U-\the\pubnum}
\date={\today}
\newcommand{\frontpageskip}{\vspace{12pt plus .5fil minus 2pt}}
\renewcommand{\title}[1]{\frontpageskip
	\begin{center}{\titlefont #1}\end{center}\par}
\renewcommand{\author}[1]{\frontpageskip\par\begin{center}
	{\authorfont #1}\end{center}
	%\par
	\nobreak
	}
\newcommand{\andauthor}{\frontpageskip\centerline{and}\author}
\newcommand{\authors}{\frontpageskip\noindent}
\newcommand{\address}[1]{\par\begin{center}{\sl #1}\end{center}\par}
\newcommand{\andaddress}{\par\centerline{\sl and}\address}
\renewcommand{\thanks}[1]{\footnote{#1}}
\renewcommand{\abstract}{\par\frontpageskip\centerline{\fourteencp Abstract}
	\vspace{8pt plus 3pt minus 3pt}}
\newcommand\YITP{\address{Uji Research Center, 
	       Yukawa Institute for Theoretical Physics\\
               Kyoto University,~Uji 611,~Japan\\}}
\thispagestyle{empty}
%%%%%%%%%%%%%%%%%%%%% end of YITP/U defs. %%%%%%%%%%%%%%%%%%%%%%%
%
%\nopubblock  % delete '%' in making submit-version 
\pubnum{94-9 \cr UTAP-179}
\date{April, 1994}
\titlepage

\title{Numerical analysis of the dynamics \\ 
of a cosmic string loop as a vortex}

\author{\sc Michiyasu Nagasawa}
\address{Department of Physics, School of Science,  \\
The University of Tokyo, Tokyo 113, Japan}

\andauthor{\sc Jun'ichi Yokoyama}
\YITP

\abstract{
Time evolution of a circular cosmic string loop is investigated by
numerically solving the field equations for the scalar and the gauge
fields consisting of the vortex.  It is shown that the result agrees
with an analytic estimate based on the Nambu-Goto action, which
supports its validity in analyzing nonstraight and rapidly moving strings.
}

\newpage

\newcommand{\gsim}{\mbox{\raisebox{-1.0ex}{$\stackrel{\textstyle >}
{\textstyle \sim}$ }}}
\newcommand{\lsim}{\mbox{\raisebox{-1.0ex}{$\stackrel{\textstyle <}
{\textstyle \sim}$ }}}
\renewcommand{\baselinestretch}{1.7}
\newcommand{\beq}{\begin{equation}}
\newcommand{\eeq}{\end{equation}}
\newcommand{\beqa}{\begin{eqnarray}}
\newcommand{\eeqa}{\end{eqnarray}}
\newcommand{\lmk}{\left(}
\newcommand{\rmk}{\right)}
\newcommand{\lkk}{\left[}
\newcommand{\rkk}{\right]}
\newcommand{\bfx}{{\bf x}}
\newcommand{\bfk}{{\bf k}}
\newcommand{\gtilde}
{~ \raisebox{-1ex}{$\stackrel{\textstyle >}{\sim}$} ~}
\newcommand{\ltilde}
{~ \raisebox{-1ex}{$\stackrel{\textstyle <}{\sim}$} ~}
\newcommand{\ba}{{\bf \phi}}
\newpage
\baselineskip 0.2cm

The cosmic string scenario is one of the major proposed formation
mechanisms of  density fluctuations in the universe \cite{cst}.
Although it is true that quantum fluctuations produced during
inflation provide realization of primordial density
fluctuations more economically \cite{influc}, 
cosmic string scenario is no less important in the
present situation of cosmology that the type of the dark matter
dominating mass density of the universe is yet unknown, because
the latter can induce structure formation even in 
the hot-dark-matter-dominated universe in which linear density 
fluctuations on small scales are washed
away due to free streaming.  Note also that, contrary to the usual prejudice,
it is easy to have cosmic strings compatible with inflation
\cite{stinf}, which is
anyway indispensable to realize globally homogeneous and nearly flat
spacetime apart from the origin of fluctuations \cite{oriinf}.

The most remarkable feature of the cosmic string scenario is that it
has essentially only one parameter, namely, the line density of a
string, $\mu$.  It should satisfy $G\mu \simeq 10^{-6}$ in order to
meet various observational requirements of large-scale structures
\cite{gmu}. 
On the other hand, it also suffers from an upperbound imposed by the
timing analysis of a millisecond pulsar \cite{timing}, since the stochastic
gravitational background radiation emitted by string loops can disturb
it \cite{gw}.

In the usual treatment of cosmic strings, their motion is described by
the Nambu-Goto action, which is proportional to the surface area of
the string world sheet \cite{NG}.   
Using the two dimensional metric tensor on the surface, 
\beq
h_{ab}=g_{\mu\nu}\frac{\partial X^\mu}{\partial \zeta^a}
\frac{\partial X^\nu}{\partial \zeta^b},~~~~\mu, \nu=0,1,2,3,~~~a,b=0,1,
\eeq
where $X^{\mu}(\zeta^a)$ is the spacetime trajectory of a string parametrized
by a timelike parameter $\zeta^0$ and a spacelike parameter $\zeta^1$,
the action is expressed as
\beq
  S=-\mu \int \sqrt{-\det h_{ab}}d\zeta^0 d\zeta^1.   \label{nambu}
\eeq
The above expression can be 
obtained for the Nielsen-Olesen vortex
line treating it infinitely thin and assuming the Lorentz invariance
along the string \cite{NO}.  
Strictly speaking, therefore, the Nambu-Goto action is
justified only for static and infinitely straight strings.
The equations of motion deduced from the action would adequately
describe dynamics of long strings which induce structure formation
because their curvature scale is many orders larger than their
thickness or the Compton wavelengths of the scalar and the gauge fields
consisting of the vortex line.  However, they may not correctly describe
the motion of small loops, especially near cusp-forming regions, where
velocity of the string reaches that of light.

Two approaches have been proposed so far to improve description of
string motion.
One is to calculate finite-width corrections to the Nambu-Goto action
with respect to the ratio of the string width to the curvature radius
\cite{MT}.
The other is to employ a new phenomenological energy-momentum tensor
for a string loop in which the line density and the string tension are
allowed to take different values depending on position and time \cite{EF}.
The latter research claims that these two quantities take considerably
different values near a cusp forming region and that in some cases the
line density vanishes at a cusp, which may imply snapping of a string
loop.  If this is the case, previous calculation of stochastic
gravitational radiation background may have to be reconsidered.

In the present paper we investigate the validity of the Nambu-Goto
action without resorting to perturbative or phenomenological methods.
That is, we reproduce a string loop in terms of scalar and gauge
fields and trace its motion by numerically solving field equations.
We compare a result of numerical simulation with an analytic
calculation obtained solving the equation of motion given by the
Nambu-Goto action.  As a tractable example for the both analyses, we
consider a circular loop. 

In order to generate a vortex solution, we employ the Abelian Higgs
model in the flat spacetime, whose Lagrangian is given by
\beqa
 {\cal L}&=& -\frac{1}{4}F_{\mu\nu}F^{\mu\nu}
 + (\partial^\mu -ieA^\mu)\phi^* (\partial_\mu + ieA_\mu)\phi
 - V[\phi], \nonumber \\
 F_{\mu\nu}&=&\partial_\mu A_\nu - \partial_\nu A_\mu, \\
 V[\phi]&=& \frac{\lambda}{2}(|\phi|^2-v^2)^2.  \nonumber 
\eeqa
The equations of motion for the complex scalar field $\phi(\bfx,t)$ and
the gauge field $A_\mu(\bfx,t)$ read,
\beqa
 \lkk \Box + 2ieA^\mu\partial_\mu - e^2A_\mu A^\mu 
+ie(\partial_\mu A^\mu)\rkk \phi + \lambda (|\phi|^2-v^2)\phi=0, \label{eqma}\\
 \partial_\mu\partial_\nu A^\nu - \Box A_\mu 
= ie(\phi^*\partial_\mu \phi - \phi\partial_\mu \phi^*)
+ 2e^2A_\mu\phi^*\phi, \label{eqmb}
\eeqa
respectively.  We take $\lambda=0.01$, $e=1$, and $v=1$ and impose the
Lorentz gauge condition, $\partial_\mu A^\mu=0$, for numerical
calculation.

In order to set up the initial configuration, using the relaxation 
method\cite{NR}, we first reproduce the
static and infinitely straight Nielsen-Olesen vortex \cite{NO}, in which
amplitudes of the scalar field and the gauge field are given as a
function of the distance from the string core.
Then it is bent artificially to make a circular loop.  
To be more specific, we calculate the minimum distance to the
trajectory of the string core at each point and assign the
corresponding field amplitudes just as the Nielsen-Olesen vortex,
while varying the phase of the scalar field uniformly
and the direction of the gauge field correspondingly.
Since we also set $\dot{\phi}=0$ and $\dot{A}_\mu=0$ initially,
where an overdot denotes time derivation, 
the second time derivative of both $\phi$ and $A_\mu$ is nonvanishing
then due to bending.  But it is small because we take the initial
radius of a loop large compared with the core radius of a string.

Making use of the rotational symmetry, the above field configuration 
is embedded in the center of a square
of $200^2$ meshes perpendicular to the string loop with the boundary
condition $\partial_i A^\mu =0$ and $\partial_i\phi=0$ where
$\partial_i$ denotes spatial derivatives.  In fact only a quarter of 
the square is required for numerical computation due to the reflective
symmetry.
We take the core radius or $\frac{1}{\sqrt{\lambda}v}$ equal to
$10\Delta x$ and the initial radius of the loop to be $50\Delta x$
with $\Delta x$ being the mesh size.  Then time evolution of the loop is
traced solving equations.\ (\ref{eqma}) and (\ref{eqmb}) in terms of the leap
frog method with a constant time step $\Delta t=0.1\Delta x$.

On the other hand, taking a cylindrical coordinate
$ds^2=-dt^2+dr^2+r^2d\theta^2+dz^2$ and $\zeta^0=t$ and $\zeta^1=\theta$,
the Nambu-Goto action (\ref{nambu}) 
for a circular loop is given by
\beq
  S= -2\pi\mu \int R(t)\sqrt{1-\dot{R}^2(t)} dt,
\eeq
where $R(t)$ is the radius of the loop.  Then the equation of motion
reads
\beq
  \ddot{R}(t)R(t)-\dot{R}^2(t)+1=0,
\eeq
which is easily solved as
\beq
  R(t)=R_0\left | \cos\lmk\frac{t}{R_0}\rmk \right |,  \label{radius}
\eeq
where we have assumed the loop is at rest with radius $R_0$ initially
at $t=0$.  The speed of the loop motion is given by 
\beq
  |\dot{R}(t)|= \left | \sin\lmk\frac{t}{R_0}\rmk \right |,  \label{speed}
\eeq
which reaches the speed of light at $t=\frac{\pi}{2}R_0 \simeq 78.5
\Delta x$ ($R_0=50\Delta x$), when the loop shrinks to a point.

Now we are in a position to display our result of numerical
calculation and compare it with the above analytic estimate (\ref{radius}).
Figure 1 depicts time evolution of the potential energy density,
$V[\phi]$, of the scalar field up to the stage the loop shrinks completely.
Due to the initial finite second derivative, the string loop
is accelerated and its radius approaches zero.
Figure 2 shows evolution of scalar field amplitude in
terms of $F(\bfx, t)\equiv 1-|\phi(\bfx,t)|^2$, which vanishes at the
potential minima and is equal to unity at the string core.
As is seen there the field configuration starts to be disturbed around
$t \simeq 60$ and after the loop has shrunken to a point, it
dissipates all the energy into scalar and vector waves without
bouncing, which is seen in the further simulation.
This is quite similar to the fate of a cylindrical domain
wall consisting of a real scalar field \cite{W}.

Finally we compare time evolution of the trajectory of the string core
$\phi(\bfx,t)=0$ with the analytic result (\ref{radius}), which is
summarized in Table 1.  As is seen there the result based on the
Nambu-Goto action agrees with the numerical calculation remarkably
well up to the stage that the loop radius becomes as small as the core
radius and that the speed of string motion is as large as $99\%$ of
that of light.  We can interpret this result as a strong support to
conventional analyses based on the Nambu-Goto action.  We also note
that the phenomenon that the string loop dissipates all the energy
without bouncing has occurred simply because we have employed a too
idealized configuration of a circular loop.  It could be avoided if
we would consider non-selfintersecting loops.  Unfortunately, however, it
is a formidable task to reproduce such a loop in terms of the scalar
and the gauge fields themselves.

In summary, we have calculated time evolution of a circular cosmic
string loop by directly solving field equations for the scalar and the
gauge fields.  The result agrees well with the analytic estimate based
on the Nambu-Goto action, so that we may conclude that the action is
applicable even for microscopically nonstraight strings moving with a
relativistic speed.
 
\vskip 1cm
\noindent 
{\Large\bf Acknowledgments}
\vskip 0.5cm
This research was supported in part by the Japanese Grant in Aid
for Science Research Fund of the Ministry of Education, Science and Culture
Nos. 3253(MN), 05218206(JY), and 05740276(JY).
Numerical computations were executed by SUN SPARC stations
at Uji Research Center,
Yukawa Institute for Theoretical Physics, Kyoto University.

\vspace{2cm}

\noindent
{\Large\bf Figure Captions}
\begin{description}
\item[Fig.\ 1] $~$ Time evolution of $V[\phi]$ on the $100^2$
 2-dimensional slice of the string loop whose radius is equal to $50\Delta x$
 initially. The upper right corner corresponds to the center of the loop and
 $r$ depicts the radial direction of cylindrical coordinates.
 These figures shows the epochs $t=0,30,60,70$ respectively.
\item[Fig.\ 2] $~$ Time evolution of $F(\bfx, t)$ on the same slice
 for the same string as Fig.\ 1 at $t=0,30,60,70,120,160$.
\end{description}

\begin{center}
{\large\bf Table 1} \\
\vspace{0.8cm}
\begin{tabular}{|c|c|c|c|}
\hline
& \multicolumn{2}{c|}{Loop Radius}& \rule{0mm}{8mm}  \\ \cline{2-3}
~~Time~~~ & ~Numerical~ & ~Analytic~ & ~~Speed~~\rule{0mm}{8mm}
\\ \hline
 0 & 50 & 50 & 0    \rule{0mm}{8mm} \\ \hline
10 & 49 & 49 & 0.199\rule{0mm}{8mm} \\ \hline
20 & 44 & 44 & 0.389\rule{0mm}{8mm} \\ \hline
30 & 41 & 41 & 0.565\rule{0mm}{8mm} \\ \hline
40 & 34 & 34 & 0.717\rule{0mm}{8mm} \\ \hline
50 & 27 & 27 & 0.841\rule{0mm}{8mm} \\ \hline
60 & 18 & 18 & 0.932\rule{0mm}{8mm} \\ \hline
70 &  8 & 8.5& 0.985\rule{0mm}{8mm} \\ \hline
75 &  2 & 3.5& 0.997\rule{0mm}{8mm} \\ \hline
\end{tabular} 
\end{center}
\vspace{0.8cm}

Time evolution of the loop radius in unit of $\Delta x$.  The third
column shows analytic estimate based on Eq.\ (\ref{radius}) and the
fourth the speed of the loop motion (\ref{speed}).


\newpage
\begin{thebibliography}{30}
\bibitem{cst}
  Ya.B.\ Zel'dovich, Mon.\ Not.\ R.\ astr.\ Soc. {\bf 192}(1980)663. \\*
  A.\ Vilenkin, Phys.\ Rev.\ Lett. {\bf 46}(1981)1169. \\*
  A.\ Vilenkin, Phys.\ Rep. {\bf 121}(1985)263.\\*
  G.\ Gibbons, S.\ Hawking, and T.\ Vachaspati eds. 
  {\it The formation and evolution of cosmic strings}
  (Cambridge University Press, Cambridge, 1990).
\bibitem{influc}
   S.W.\ Hawking, Phys.\ Lett. {\bf B115}(1982)295.\\*
   A.A.\ Starobinsky, Phys.\ Lett. {\bf B117}(1982)175.\\*
   A.H.\ Guth and S-Y.\ Pi, Phys.\ Rev.\ Lett. {\bf 49}(1982)1110.
\bibitem{stinf}
   J.\ Yokoyama, Phys.\ Lett. {\bf B212}(1988)273.\\*
   M.\ Nagasawa and J.\ Yokoyama Nucl.\ Phys. {\bf B370}(1992)472.
\bibitem{oriinf}
   A.H.\ Guth, Phys.\ Rev. {\bf D23}(1981)347.\\* 
   K.\ Sato, Mon.\ Not.\ R.\ astr.\ Soc. {\bf 195}(1981)467. 
\bibitem{gmu}
   J.\ Silk, and A.\ Vilenkin, Phys.\ Rev.\ Lett.\ 
{\bf 53}(1984)1700.\\*
   T.\ Vachaspati and A.\ Vilenkin, Phys.\ Rev.\ Lett.\ {\bf 67}(1991)1057.
\bibitem{timing}
   D.R.\ Stinebring, M.F.\ Ryba, J.H.\ Taylor, and R.W.\ Romani, 
   Phys.\ Rev.\ Lett. {\bf 65}(1990)285.
\bibitem{gw}
   R.R.\ Caldwell and B.\ Allen, Phys.\ Rev. {\bf D45}(1992)3447 and
references cited therein.
\bibitem{NG}
   Y.\ Nambu, Proc.\ Int.\ Conference on Symmetries and Quark Models,
(Wayne State University, 1969).\\*
   T.\ Goto, Prog.\ Theor.\ Phys. {\bf 46}(1971)1560.
\bibitem{NO}
   H.B.\ Nielsen and P.\ Olesen, Nucl.\ Phys. {\bf B61}(1973)45. 
\bibitem{MT}
   K.\ Maeda and N.\ Turok, Phys.\ Lett.\ B {\bf 202}(1988)376. \\*
   R.\ Gregory, Phys.\ Lett.\ B {\bf 206}(1988)199. 
\bibitem{EF}  
   Y.\ Eriguchi and T.\ Futamase, Prog.\ Theor.\ Phys. {\bf 86}(1991)1227.
\bibitem{NR}
   W.H.\ Press, B.P.\ Flannery, S.A.\ Teukolsky and W.T.\ Vetterling,
   {\it Numerical Recipes} (Cambridge University Press, Cambridge, 1986).
\bibitem{W}
   L.M.\ Widrow, Phys.\ Rev. {\bf D40}(1989)1002.
\end{thebibliography}
\end{document}